\documentclass[%
 aip,
 amsmath,amssymb,
 reprint,%
]{revtex4-1}

\usepackage{graphicx}
\usepackage{dcolumn}
\usepackage{bm}
\usepackage[utf8]{inputenc}
\usepackage[T1]{fontenc}
\usepackage{verbatim}
\usepackage{mathptmx}
\usepackage{etoolbox}
\usepackage{subcaption}
\usepackage{siunitx}
\usepackage{lipsum}
\usepackage{hyperref}
\hypersetup{
    colorlinks=true,
    linkcolor=blue,
    citecolor=blue,
    filecolor=magenta,      
    urlcolor=blue,
    pdfpagemode=FullScreen,
    }
\usepackage{cleveref}

\newcommand{\grillix}{\texttt{GRILLIX}}
\newcommand{\genex}{\texttt{GENE-X}}

\makeatletter
\def\@email#1#2{%
 \endgroup
 \patchcmd{\titleblock@produce}
  {\frontmatter@RRAPformat}
  {\frontmatter@RRAPformat{\produce@RRAP{*#1\href{mailto:#2}{#2}}}\frontmatter@RRAPformat}
  {}{}
}%
\makeatother
\begin{document}

\preprint{AIP/123-QED}

\title{On the mechanism of Pedestal Relaxation Events - \\ Insights gained by turbulence simulations with GRILLIX}
\author{Christoph Pitzal}%
\email{christoph.pitzal@ipp.mpg.de}
\author{Andreas Stegmeir}%
\author{Tim Happel}
\author{Kaiyu Zhang}
\author{Konrad Eder}
\author{Wladimir Zholobenko}
\author{Philipp Ulbl}
\author{Manuel Herschel}
\author{Frank Jenko}
\author{The ASDEX Upgrade Team}
\affiliation{ 
Max Planck Institute for Plasma Physics, Boltzmannstr. 2, 85748 Garching, Germany
}%

\date{\today}

\begin{abstract}

Pedestal Relaxation Events (PREs) appear in I-mode discharges close to the I-H transition. Although they show certain similarities with Edge Localised Modes (ELMs), i.e. periodic energy ejections, the underlying mechanism seems to be very different from the mechanism responsible for ELMs. In this manuscript, we present global trans-collisional fluid simulations of an I-mode discharge in ASDEX Upgrade using \grillix. We observe multiple PREs during the simulation, which reproduce a range of experimentally observed PRE characteristics. Furthermore, a detailed analysis of various mode properties in our simulation allows us to pinpoint the underlying mechanism responsible for triggering PREs to Micro-Tearing Modes (MTMs). The system is analysed dynamically by evaluating density and electron temperature gradient lengths at the OMP position, where the MTM is located and grows over time. The path taken by the system in gradient length space is compared to a growth-rate estimate calculated by linear theory in simplified slab geometry, providing excellent agreement. Building on these insights, we sketch a qualitative picture of a full PRE cycle.
Finally, we discuss the influence of the recently implemented Landau-fluid closure and the challenges of simulating low collisionality regimes with trans-collisional fluid models, like the one employed by \grillix.

\end{abstract}

\maketitle

\section{\label{sec:Intro}Introduction}

Operational regimes for future fusion reactors should provide high confinement, while not exceeding material limits at the wall of the device. Studying and gaining deeper insights into such reactor-relevant regimes is therefore a task of major importance on the way to the first fusion power plant. The improved energy confinement mode (I-mode) is such a reactor-relevant regime, which is characterised by elevated energy confinement time, low impurity content and the absence of Type-I Edge Localised Modes (ELMs) \cite{Whyte_2010, Happel_2017}. Although there are no Type-I ELMs present in I-mode, some discharges show Pedestal Relaxation Events (PREs), especially when the operating point is close to the I-H transition. PREs cause periodic energy ejections on significantly lower levels than ELMs, about 1\% of the plasma stored energy \cite{Silvagni_2020}. In contrast to ELMs, which are triggered by Peeling-Ballooning modes that become unstable in the pedestal region \cite{huijsmans_2015}, PREs are expected to be caused by a different mechanism. PREs were observed and studied in ASDEX Upgrade and Alcator C-Mod \cite{Silvagni_2022}, a similar phenomenon was also observed in EAST \cite{Zhong_2022}. In experiments, PREs affect mainly the electron temperature profile and show magnetic activity during the event \cite{Silvagni_2020, Silvagni_2022}.
Attempts to simulate PREs with a gyrofluid model have been performed \cite{Manz_PRE_2021}. The authors expected Micro-Tearing Modes (MTMs) as presumable candidates to cause PREs. However, within their qualitative simulations in simple geometry, they observed for the PRE a transition from drift-wave to interchange-like turbulence with no significant electromagnetic transport.

Simulating edge conditions in I-mode with fluid models is rather challenging due to the low collisionality present in the edge region of a fusion device ($\rho_{\mathrm{pol}} > 0.9$). The edge turbulence fluid code \grillix \cite{stegmeir_grillix19} was recently extended in the direction of low collisionality by implementing a Landau-fluid closure for parallel conductive heat fluxes \cite{pitzal_landau_2023}. 

\begin{figure*}[t]
    \centering
    \includegraphics[width=0.97\linewidth]{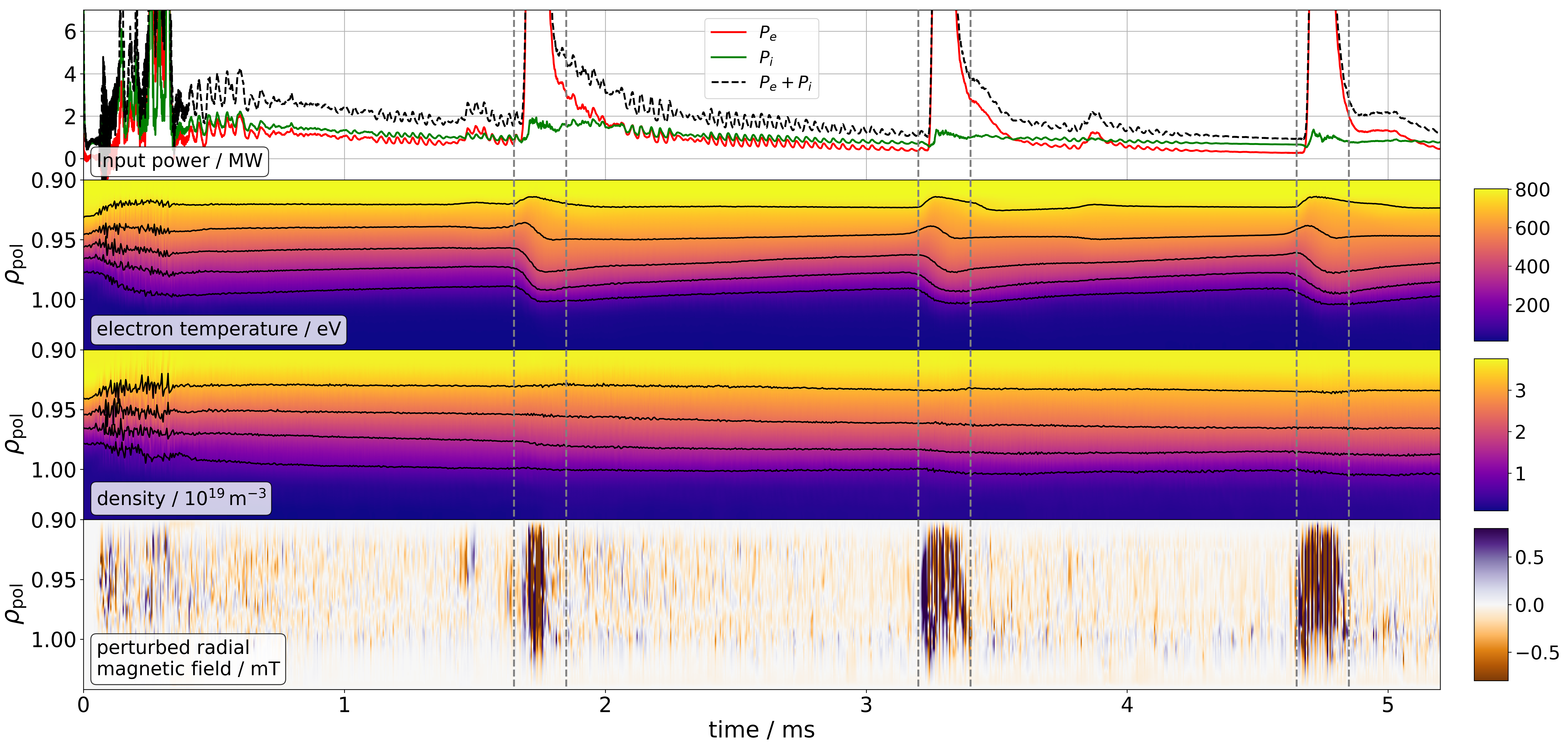}
    \caption{Time traces of the input power and OMP profiles of electron temperature, density and perturbed radial magnetic field. The PREs are marked with grey dotted lines.}
    \label{fig:timetrace_power_omp_quant}
\end{figure*}

Within this paper, \grillix~is used for studying I-mode conditions, the relevant parameters and geometry are taken from the ASDEX Upgrade discharge \# 37980, a steady I-mode discharge with no PREs present at the relevant time. We present a simulation spanning a physical time of more than $5 \; \mathrm{ms}$, where three intermittent bursts occur, that are identified as PREs. The focus of this work is on a qualitative picture of PREs. The characteristics of the PREs in the simulation are compared to experimental findings. A detailed analysis of the underlying microinstability reveals that the PREs in our simulation are caused by MTMs. The clearest fingerprints are the visible tearing parity, the radial electromagnetic electron heat transport and the dispersion relation computed from the simulation data, which matches nicely linear kinetic estimates for MTMs \cite{Hatch_2021}. A distinct difference is that the stored energy after a PRE in our simulation is increased, while in the experiment, it is decreased. However, we clarify that this originates in the way boundary conditions are applied in the simulation. Furthermore, we investigate the dynamic behaviour of the system by calculating the density and temperature gradient lengths at $\rho_{\mathrm{pol}} = 0.95$, where the MTM is centred. The path taken by the system is visualised in gradient-length space and agrees very well with an analytical estimate of the MTM growth rate. This linear estimate is calculated in the present article for our fluid plasma model, including the Landau-fluid closure, in simplified slab geometry following previous theoretical work \cite{Hassam_1980}, predicting positive linear growth rates for MTMs. We investigate the impact of the Landau-fluid closure by comparison to a simulation with the Braginskii closure, including heat-flux limiters, which does not show any PREs. The linear analysis is performed as well for the Braginskii closure, revealing that, in contrast to the Landau-fluid closure, MTMs are linearly stable in this case. Finally, we discuss why PREs are observed in our simulation even though they do not occur in the discharge under investigation, and what role is played by the parallel resistivity. \\

This paper is organised as follows. In \cref{sec:MTMs}, a short introduction to MTMs is provided, \cref{sec:Turb_Sim} explains the simulation setup in detail. In \cref{sec:PREs_caused_by_MTMs} we elaborate on the PREs present in our simulations and provide evidence that the presented PREs are caused by MTMs. In \cref{sec:Global_PRE_mechanism}, we propose a global mechanism for the steps of a PRE cycle. The importance of the fluid closure used for the simulations is highlighted in \cref{sec:Fluid_closure}, including the linear analysis for the growth rate of our fluid model. Finally, in \cref{sec:Discussion} a discussion and in \cref{sec:Summary_and_Outlook} a summary and outlook of this work are provided.

\section{\label{sec:MTMs}Microtearing Modes}

MTMs are electromagnetic micro-instabilities with high poloidal mode number $m$. For most experimentally relevant conditions, they are driven by the electron temperature gradient \cite{Hazeltine_1975, Gladd_1980, Hatch_2021}.
The mechanism which increases the radial transport can be depicted as follows: MTMs form magnetic islands within the plasma. When the islands grow large enough, the magnetic field becomes stochastic, parallel and perpendicular transport couple and the radial transport is increased \cite{Guttenfelder_PRL_2011, Doerk_PRL_2011}. With this picture, it becomes evident that MTMs feature increased electromagnetic transport, due to the field stochastisation, and predominantly an increase in the radial electron heat transport, due to the rapid parallel electron heat conduction. 

A simple kinetic dispersion relation for the complex frequency $\omega$ of an MTM can be found in \cite{Hatch_2021,Hazeltine_1975} and reads

\begin{equation}
    \left( \nu - 0.51 i \omega \right) \left( \omega - \omega_{p}^{\star} \right) - 0.8 \omega_{T}^{\star} \nu = 0
    \label{eq:disp_rel_hatch}
\end{equation}

with $\omega_{p}^{\star} = k_y \rho_s c_s \left( 1 / L_n + 1 / L_{Te} \right)$ and $\omega_{T}^{\star} = k_y \rho_s c_s \left( 1 / L_{Te} \right)$, $\nu$ the electron collision frequency \cite{huba1998nrl}, $i$ the imaginary unit, $k_y$ the binormal wavenumber, $\rho_s = \sqrt{T_e M_i} / (e B)$ the sound Larmor radius, $c_s = \sqrt{T_e / M_i}$ the ion sound velocity. The gradient lengths are defined as $L_n = -n / (\nabla n)$ and $L_{Te} = -T_e / (\nabla T_e)$. From this dispersion relation, it can be deduced that the real propagation frequency $\omega_r = \mathfrak{Re}(\omega)$ of an MTM is $\omega_r \approx \omega_p^{\star}$. The propagation direction is in electron diamagnetic direction.

Previous theoretical work on fluid models revealed some mechanisms that are able to recover MTMs, when included in a plasma fluid turbulence model, e.g. a time-dependent thermal force \cite{Hassam_1980} or the interaction between trapped and passing particles \cite{Guttenfelder_PRL_2011}.
The effect present in our model is the Landau-fluid closure \cite{pitzal_landau_2023}, which is similar to the effect of a time-dependent thermal force, since it also alters the parallel heat conduction and introduces non-local effects to the fluid model.

\section{\label{sec:Turb_Sim}Turbulence simulations}

\begin{figure}[t]
    \centering
    \includegraphics[width=\linewidth]{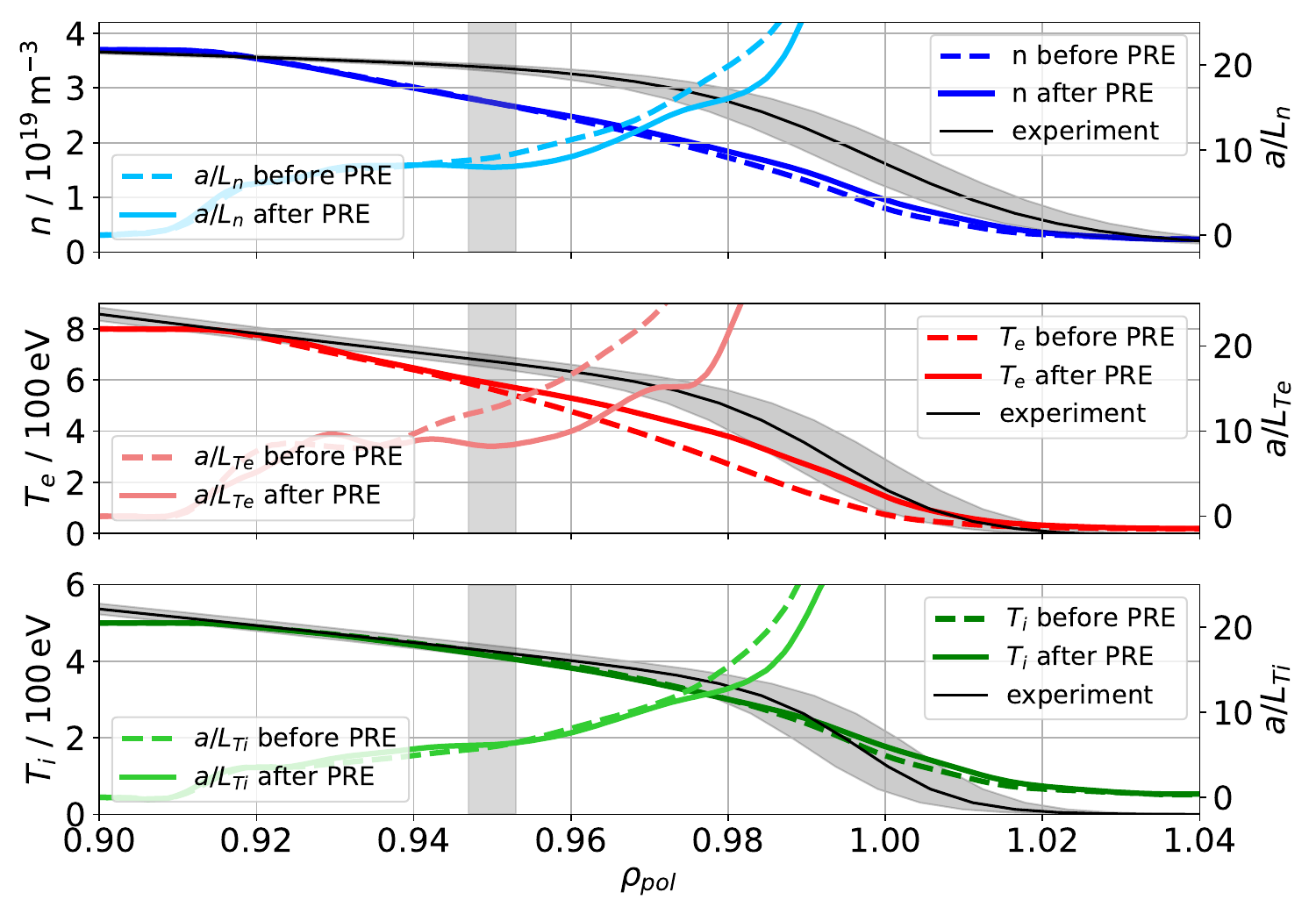}
    \caption{Density, electron and ion temperature profiles before and after the second PRE, exact time windows are $\Delta t_{\mathrm{before}} = 3.10-$ $3.20 \; \mathrm{ms}$ and $\Delta t_{\mathrm{after}} = 3.57-$ $3.67 \; \mathrm{ms}$. The area marked in grey indicates where the MTM is centred.}
    \label{fig:profiles_pre_and_post_PRE}
\end{figure}

The simulation setup is based on the ASDEX Upgrade (AUG) discharge \# 37980, an I-mode discharge performed in unfavourable configuration (upper-single null). For the simulation, the time point $t = 4.1 \, \mathrm{s}$ is considered, where the discharge featured a stationary I-mode. The magnetic field was $B = 2.5 \, \mathrm{T}$ on axis. The plasma was heated by electron-cyclotron-resonance heating (ECRH) and by neutral-beam injection (NBI), after compensation for the radiation losses $P_{\mathrm{rad}}$, which are not modelled by \grillix, the input power was $1.9 \, \mathrm{MW}$.
For the simulation, 16 poloidal planes were used. A constant grid distance of $1.45 \, \mathrm{mm}$ within a plane is used. The inner core boundary values of $n(\rho_{\mathrm{pol}} = 0.91) = 3.8 \cdot 10^{19} \, \mathrm{m^{-3}}$, $T_e(\rho_{\mathrm{pol}} = 0.91) = 800 \, \mathrm{eV}$ and $T_i(\rho_{\mathrm{pol}} = 0.91) = 500 \, \mathrm{eV}$ are taken from the experiment, see \cref{fig:profiles_pre_and_post_PRE}. These boundary conditions are imposed by adaptive sources, which are active between $\rho_{\mathrm{pol}} = 0.90-0.91$ and inject heat and particles to keep the values of $n$, $T_e$ and $T_i$ close to the target value. The input of heating power and particles is recorded during the simulation. The boundary conditions at the divertor target are sheath boundary conditions with sheath transmission factors of $\gamma_{sh,e} = 1.0$ and $\gamma_{sh,i} = 0.1$. For a more detailed description of the boundary conditions, we refer to \cite{pitzal_landau_2023}.
The presented simulation is performed employing the same physical model as in \cite{pitzal_landau_2023}, i.e. a drift-reduced electromagnetic Braginskii model including trans-collisional extensions, namely a neoclassical extension of the ion viscosity and a Landau-fluid closure for parallel heat fluxes. Instead of the one-moment neutral gas model, which was used previously, we use here a three-moment model \cite{Eder_2025}. 


To provide an overview of the simulation, we plot a time trace of the input power of the adaptive source near the core boundary and outboard midplane (OMP) profiles of electron temperature, density and perturbed radial magnetic field in \cref{fig:timetrace_power_omp_quant}. Herein, we find intermittent bursts marked by grey dotted lines, where we observe a significant increase in heating power, mainly in electron input power $P_e$. The scale of the input power is cut for visual reasons. The peak heating power for electrons in the first peak is $32 \, \mathrm{MW}$, in the second peak $23 \, \mathrm{MW}$ and in the third peak $22 \, \mathrm{MW}$. A simulation with $50 \%$ higher resolution in all spatial directions was run until the first PRE, which is in line with the investigated standard resolution simulation. Furthermore, we see strong magnetic activity in the radial perturbed magnetic field and a substantial effect on the electron temperature profile, while the density profile is nearly unaffected. 
In the following, we will substantiate our claim that these three events are indeed PREs, which manifest in the simulation as well as in the experiment as periodically occurring and short-lasting increases of radial heat transport, which share qualitatively the underlying physical mechanism.  

The characteristics of the three intermittent events in our simulation align nicely with the properties of experimentally observed PREs. More specifically, for PREs observed in experiments, mainly the electron temperature profile flattens rather than the density profile \cite{Silvagni_2020}. For the ion temperature profile, no measurements were available, unfortunately. Furthermore, during PREs, there is strong magnetic activity observed \cite{Silvagni_2022,Manz_PRE_2021}.
More importantly, the duration of a PRE of ca. $0.3 \; \mathrm{ms}$ is very close to the experimentally observed duration \cite{Silvagni_2020}.
To show the effect of the PREs more quantitatively, we plot density, electron and ion temperature profiles plus their gradient lengths before and after the second PRE in \cref{fig:profiles_pre_and_post_PRE}. Again, mainly the electron temperature profile is affected. Taking a close look at \cref{fig:profiles_pre_and_post_PRE}, it does not seem intuitive to call these events PREs at first, since the electron temperature is increased in the whole domain after the event. An explanation can be found by considering the adaptive sources near the core boundary. When the radial transport in the simulation domain increases, the adaptive sources try to maintain the target value by increasing the injected heating power. The significantly increased input power during the event is responsible for the increase in the stored energy. To make sure that the PREs present in our simulation are caused by internal dynamics and are not artificially induced by the boundary conditions, we investigate the causality of events. Therefore, we take a look at \cref{fig:timetrace_power_omp_quant} again. Investigating herein the second PRE, we can observe that first, the electron temperature profile starts to flatten. As soon as this perturbation reaches the location of the adaptive source, the input power for electrons $P_e$ increases drastically to keep up the core value. 
The mode, which is growing and causing an increase in radial transport, appears close to $\rho_{\mathrm{pol}} = 0.95$. At this radial location, where the PRE occurs, the electron temperature gradient relaxes, which is visible in \cref{fig:profiles_pre_and_post_PRE} for $a/L_{Te}$ before and after the PREs, just like in the experiment.  
We want to emphasize here that the PRE in the simulation is at $\rho_{\mathrm{pol}} = 0.95$, which can be considered as pedestal top, in comparison to the experiment, where the PRE is located at the maximum $T_e$ gradient region, i.e. inside the pedestal. This point will be discussed in more detail in \cref{sec:Discussion}.

\section{\label{sec:PREs_caused_by_MTMs}PREs caused by MTMs}

\begin{figure}[t]
    \centering
    \includegraphics[width=\linewidth]{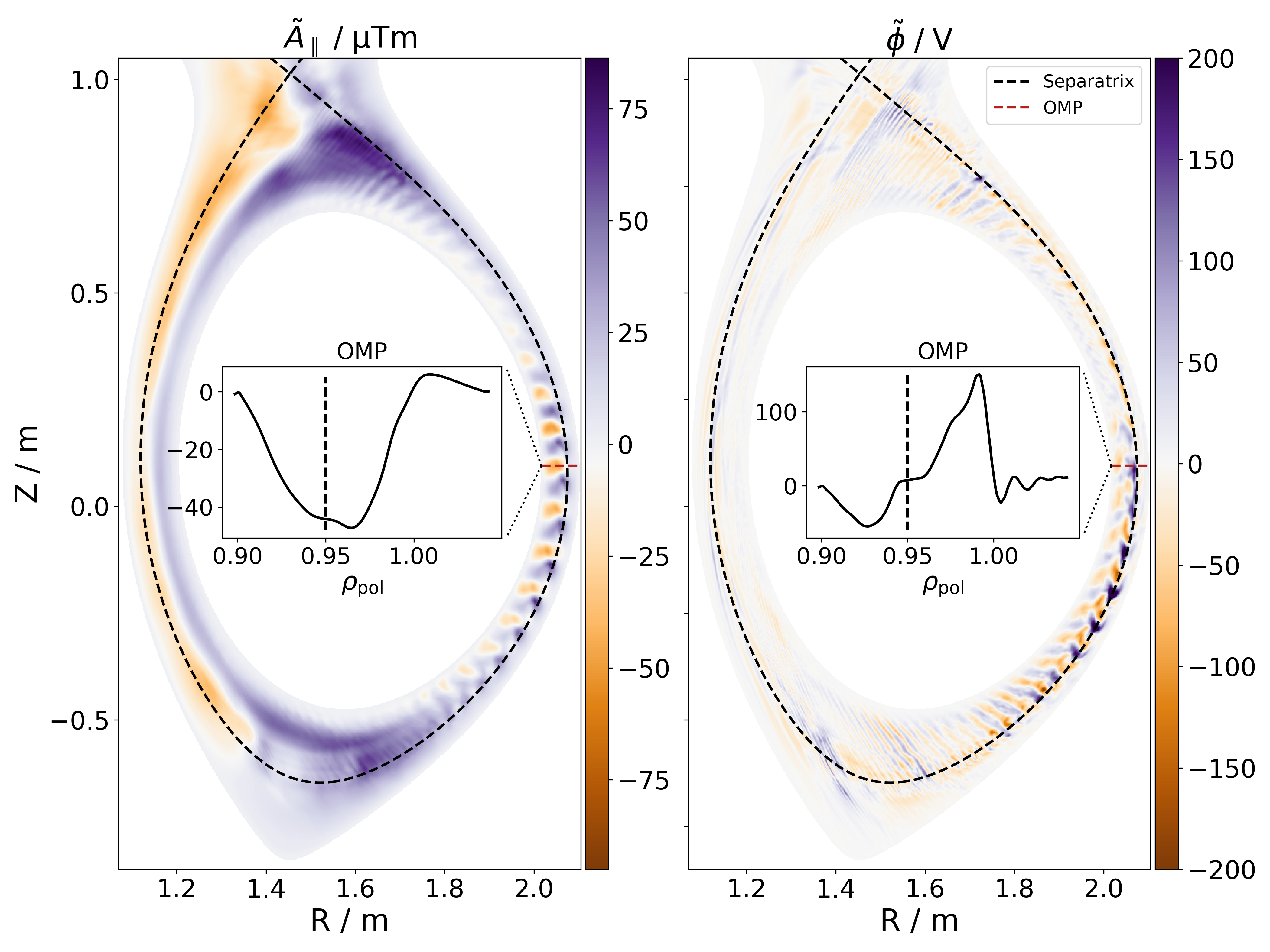}
    \caption{Snapshot of $\tilde{A}_{\parallel}$ and $\tilde{\phi}$ during the second PRE at $t=3.23 \; \mathrm{ms}$ showing tearing parity.}
    \label{fig:A_par_2D}
\end{figure}

\begin{figure}[b]
    \centering
    \includegraphics[width=\linewidth]{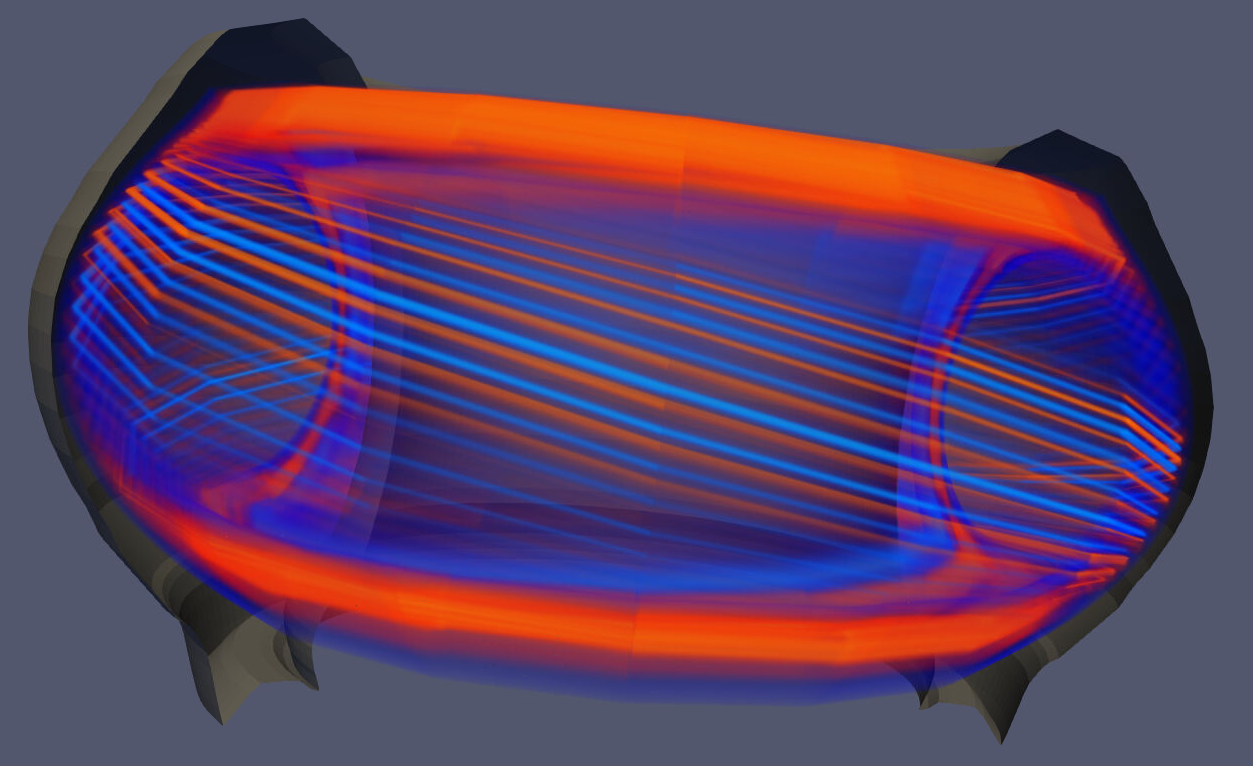}
    \caption{3D visualisation of $\tilde{A}_{\parallel}$ during the second PRE, the first wall and divertor of AUG are depicted in dark grey.}
    \label{fig:A_par_3D}
\end{figure}

After comparing PREs in simulation and experiment, this chapter aims to provide sufficient proof that the underlying microinstability responsible for the PREs in our simulation is a MTM. 
The first fingerprint was already mentioned earlier, namely, that mainly the radial electron heat transport is increased. Deeper analysis reveals that this increased radial electron heat transport is primarily electromagnetic, i.e. caused by magnetic flutter, as one would expect for MTMs.
The second hint towards MTMs is found by investigating the spatial structure of the fluctuating quantities $\tilde{A}_{\parallel}$ and $\tilde{\phi}$ during a PRE, as shown in \cref{fig:A_par_2D}. The mode visible in $\tilde{A}_{\parallel}$ at the low-field side peaks at the OMP around $\rho_{\mathrm{pol}} = 0.95$. For tearing modes, the eigenfunctions extend in the radial direction \cite{Hazeltine_1975}. In our case, we find even parity for $\tilde{A}_{\parallel}$, while the electrostatic potential $\tilde{\phi}$ shows odd parity. This combination of even parity in $\tilde{A}_{\parallel}$ and odd parity in $\tilde{\phi}$ is called tearing parity and is characteristic for tearing modes, including MTMs \cite{Hatch_2021}. The full three-dimensional structure of $\tilde{A}_{\parallel}$ is shown in \cref{fig:A_par_3D}, where the global extent of the mode is visible.

\begin{figure}[t]
    \centering
    \includegraphics[width=\linewidth]{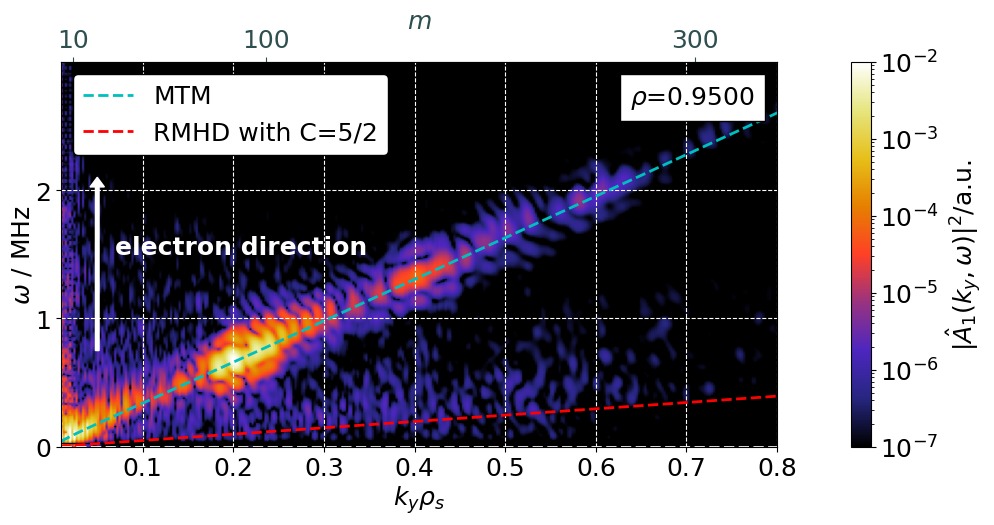}
    \caption{Dispersion Relation of $\tilde{A}_{\parallel}$ during the second burst between $t = 3.22 - 3.36 \; \mathrm{ms}$, in blue the analytical dispersion relation in \cref{eq:disp_rel_hatch}, in red an upper bound for resistive MHD \cite{Kotschenreuther_2019}.}
    \label{fig:dispersion_rel}
\end{figure}

Lastly, we want to investigate the dispersion relation, which is calculated for $\tilde{A}_{\parallel}$ on the flux surface $\rho_{\mathrm{pol}} = 0.95$ during the second PRE and is shown in \cref{fig:dispersion_rel}. This flux surface was taken, due to the location of the maximum of the mode in $\tilde{A}_{\parallel}$. In the dispersion relation, we found a mode propagating in electron diamagnetic direction in the plasma frame. This mode agrees very well with the frequency predicted by the analytical dispersion relation in \cref{eq:disp_rel_hatch}. The maximum of the mode is around $k_y \rho_s = 0.2$. When looking at the phase shifts $\alpha$ between different fluctuating quantities, we found $\alpha(\tilde{p}_e, \tilde{\phi}) \approx \alpha(\tilde{p}_i, \tilde{\phi}) \approx 0$ and $\alpha(\tilde{q}_{\parallel,e}, \tilde{A}_{\parallel}) \approx \pm \pi$, as expected for MTMs. 
Due to this list of hints and fingerprints, we are very confident in stating that the mode present in the simulation is a MTM.

\section{\label{sec:Global_PRE_mechanism}The global mechanism of PRE cycles}

\begin{figure}[t]
    \centering
    \includegraphics[width=\linewidth]{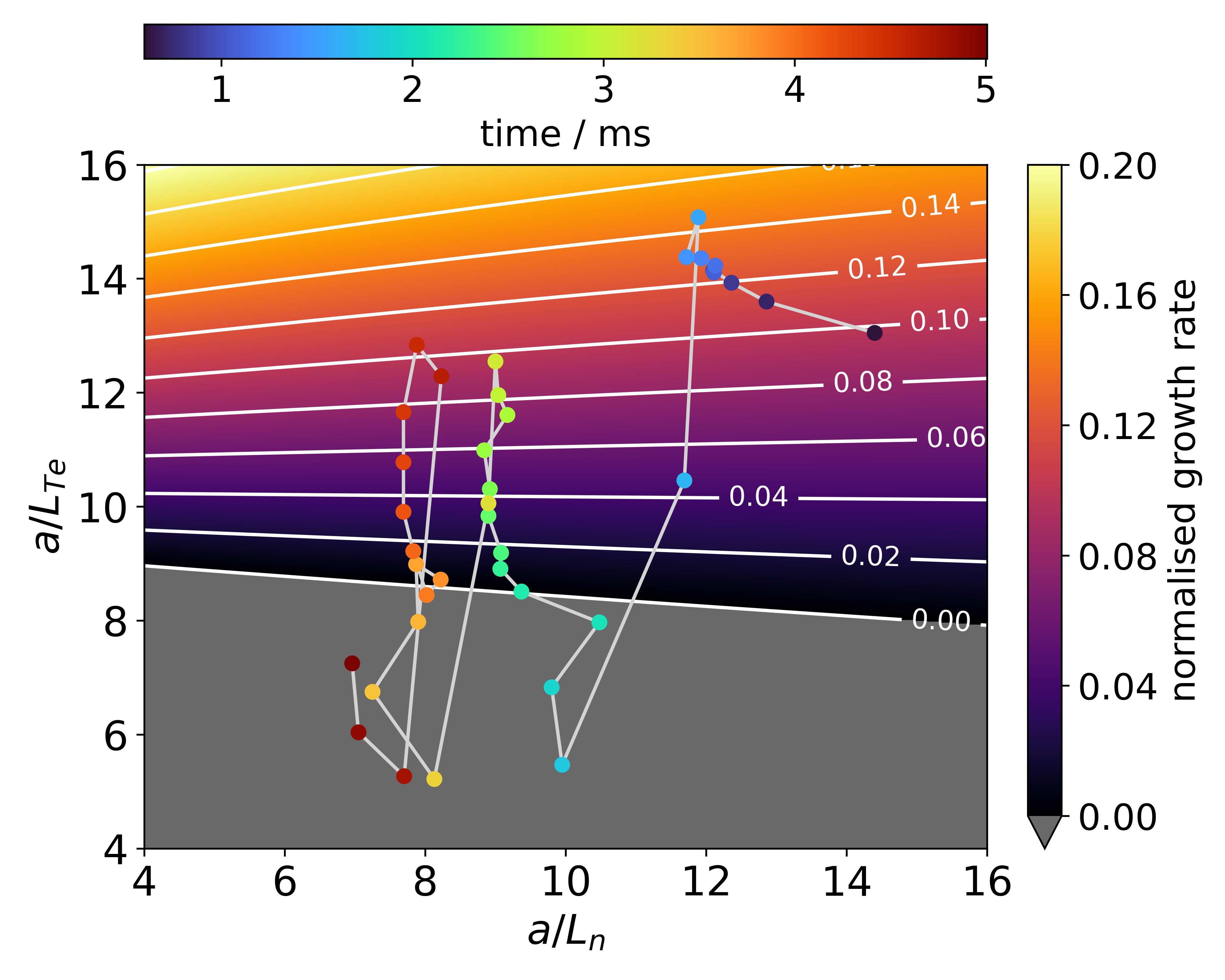}
    \caption{Path taken by the system at flux surface $\rho=0.95$ in gradient lengths space. Normalised linear growth rate $\hat{\gamma}$ from \cref{eq:lin_fluid_disp_rel} as colour plot. PREs are visible as sudden changes.}
    \label{fig:path_in_grad_length_space}
\end{figure}

\begin{figure}[b]
    \centering
    \includegraphics[width=0.95\linewidth]{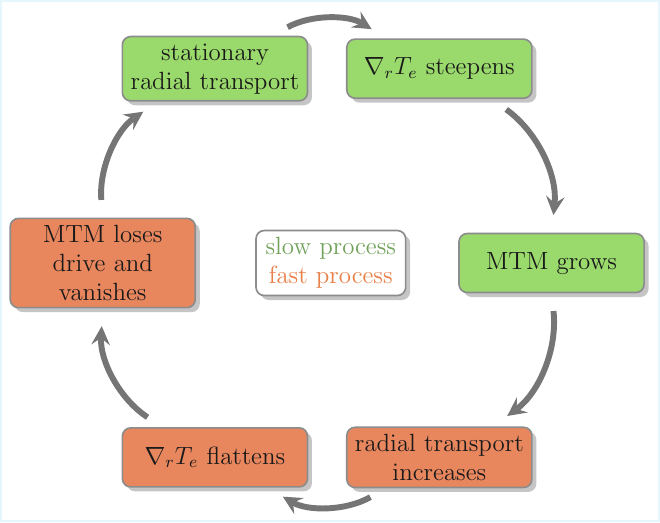}
    \caption{Sketch of PRE cycle}
    \label{fig:sketch_global_PRE_cycle}
\end{figure}

We want to investigate the dynamic process of a full PRE cycle in more detail. As mentioned in \cref{sec:MTMs}, MTMs are usually driven by the electron temperature gradient. Therefore, we perform an analysis, which traces the normalised inverse electron temperature gradient length $a / L_{Te} = - a (\nabla T_e) / T_e$ and the inverse normalised density gradient length $a / L_{n} = - a (\nabla n) / n$ at the OMP profile at the position $\rho_{\mathrm{pol}} = 0.95$ over time, with $a = 0.5 \; \mathrm{m}$ the minor radius of AUG. The early part of the simulation, where the initial instability is present and the profiles change strongly, is omitted. The result of this analysis is plotted in \cref{fig:path_in_grad_length_space}. 
The coloured background plot with contour lines shows the normalised growth rate $\hat{\gamma} = \gamma / \omega_p^{\star}$, obtained by the linear fluid dispersion relation \cref{eq:lin_fluid_disp_rel}, which is introduced in the following chapter. This linear growth rate depends on $T_e$ and $n$, which are changing slightly during the simulation. Therefore, we do not plot the linear growth rate for all points in time, but fix it shortly before the second burst. This leads to  values of $T_e = 570 \; \mathrm{eV}$ and $n = 2.7 \cdot 10^{19} \; \mathrm{m}^{-3}$. The value of $k_y \rho_s = 0.2$ was taken from the analysis in \cref{fig:dispersion_rel}. 
Negative growth rates are marked in grey in \cref{fig:path_in_grad_length_space}. According to \cref{eq:lin_fluid_disp_rel}, MTMs are linearly stable for small values of $a/L_{Te}$. 
Looking at the dynamic path of the system, marked as coloured dots, we observe that the electron temperature gradient steepens slowly over time, i.e. $\hat{\gamma}$ increases, until a MTM starts to grow. Due to the strong radial electron heat flux, the gradient flattens on a short time scale. The MTM loses its drive, as the system enters a region where MTMs are linearly stable $(\hat{\gamma} < 0)$ and vanishes eventually. This concludes one PRE cycle, and the next one begins. Within the simulation time of approx. $5 \; \mathrm{ms}$, three such events are observed in the simulation. The density gradient is slightly affected by the MTMs as well, but much weaker than $a/L_{Te}$. Overall, we observe for the density gradient a trend of relaxing slowly over the whole simulation time. There are two density sources present in our system, the adaptive source near the core, mentioned earlier, and a density source due to ionization of neutrals close to the separatrix and in the SOL. Especially the latter one takes a long time to saturate, since the ionisation source depends on the interaction between the neutral gas and plasma near the divertor. The density, which is increasing slowly near the separatrix due to ionisation, explains the slow relaxation of the density gradient in \cref{fig:path_in_grad_length_space}.  
The full chain of events leading to a PRE cycle in our simulations is summarised in \cref{fig:sketch_global_PRE_cycle}.
When investigating the linear growth rate at every point in time, we find that the last point before a PRE occurs has a value of $\hat{\gamma} \approx 0.1$ for all three PREs. Furthermore, the points after a PRE are always deep in the stable region $\hat{\gamma} < 0$.

A further point worth mentioning here is that, since MTMs are destabilized solely due to the electron temperature gradient, this insight could lead to further experimental investigations and might help to find a way of mitigating PREs.

\section{\label{sec:Fluid_closure}Importance of the employed fluid closure}

\begin{figure}[b]
    \centering
    \begin{subfigure}[b]{\linewidth}
        \centering
        \includegraphics[width=0.95\textwidth]{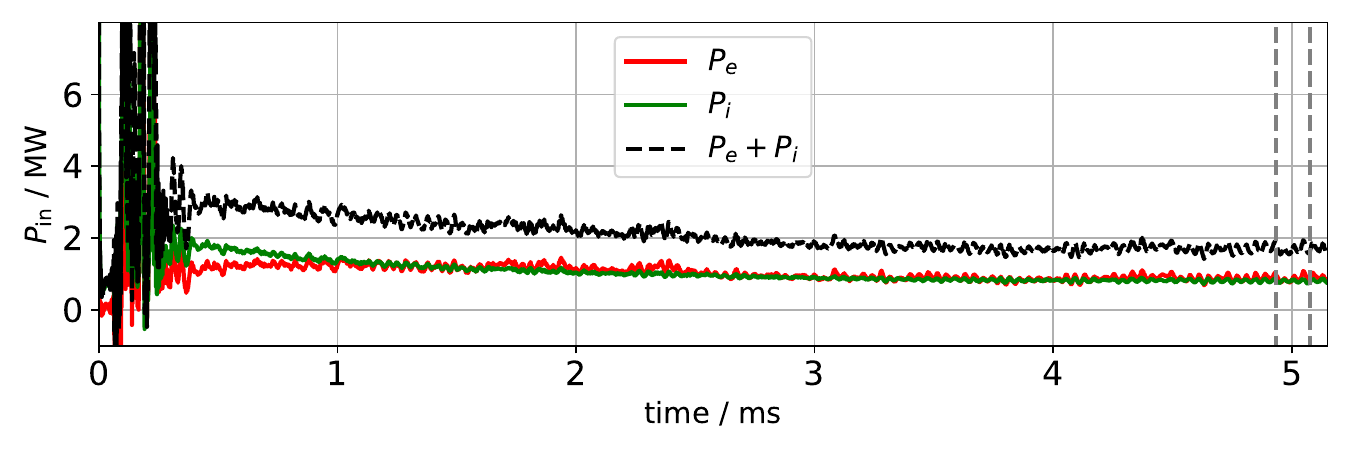}
    \end{subfigure}
    \hfill
    \begin{subfigure}[b]{0.95\linewidth}
        \centering
        \includegraphics[width=\textwidth]{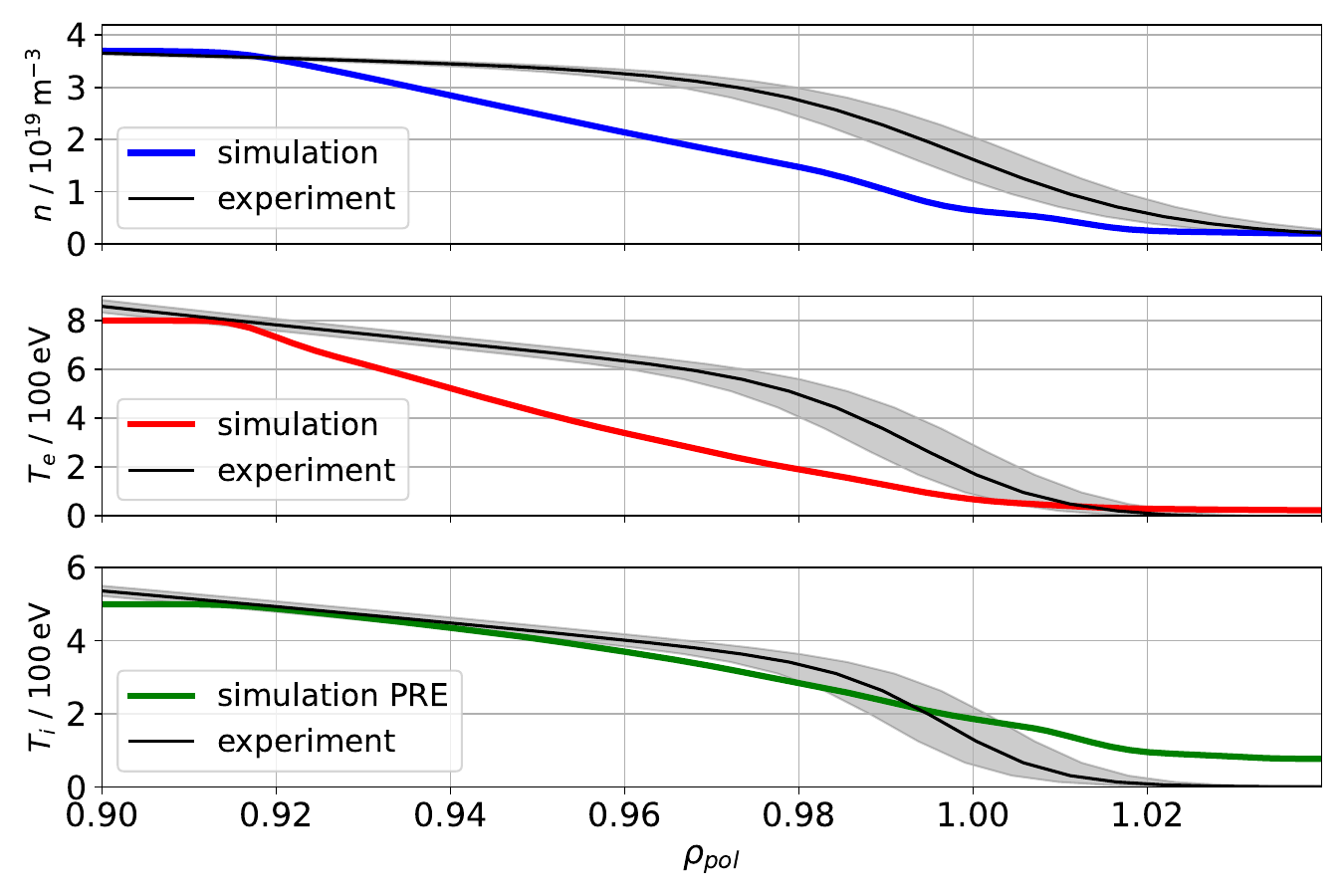}
    \end{subfigure}
    \caption{Time traces of the input power for the case with Braginskii closure with flux limiters. Below, OMP profiles of density, electron and ion temperature for the time span marked in grey dotted lines.}
    \label{fig:time_trace_and_profiles_braginskii}
\end{figure}

\subsection{Comparison to the Braginskii closure}

The presented simulation was performed with the recently implemented Landau-fluid closure for parallel heat fluxes \cite{pitzal_landau_2023}. Due to its high temperature and low density in the edge region, the I-mode scenario shows rather low collisionality in the edge region. Therefore, the Landau-fluid closure is expected to be important for investigating this regime. To test this hypothesis, the simulation presented in \cref{sec:Turb_Sim} is compared to a simulation performed with the Braginskii closure including free-streaming heat-flux limiters $(\alpha_e = \alpha_i = 1.0)$ \cite{pitzal_landau_2023}. 

A compact overview of the simulation with the Braginskii closure is shown in \cref{fig:time_trace_and_profiles_braginskii}. In contrast to the Landau-fluid simulation, we observe no intermittent events, but a very quiescent run. The associated OMP profiles for the time span marked in grey dotted lines are shown below the time traces. Especially for the electron temperature profile, no temperature pedestal is visible. In fact, these profiles would suggest an "overheated" L-mode in AUG instead of an I-mode.

\subsection{Linear analysis}

To understand this difference in more detail, we followed an analysis previously used to investigate a fluid model with a time-dependent thermal force \cite{Hassam_1980}. In the mentioned publication, further details about the setup can also be found. We calculated the linear dispersion relation for our fluid model in simplified slab geometry with a background temperature gradient in the $x$-direction, perpendicular to the direction of the magnetic field in the $z$-direction. All quantities can be split into background and fluctuating parts $f = f_0 + \tilde{f}$, where the fluctuating quantities behave as $\tilde{f} \sim \exp \left( i k_y y + i k_z z - i \omega t \right)$. Density fluctuations are neglected here, $n = n_0$. We performed this exercise for the Landau-fluid and the Braginskii closure, without heat-flux limiters. The starting point is Ohm's law 

\begin{equation}
    \eta j_{\parallel} = E_{\parallel} + \frac{1}{en} \nabla_{\parallel} p_e + 0.71 \frac{1}{e} \nabla_{\parallel} T_e \; , 
\end{equation}

with $\eta$ the resistivity, $j_{\parallel}$ the parallel current, $E_{\parallel}$ the parallel electric field, $e$ the elementary charge, $p_e$ the electron pressure, neglecting the time derivative of $j_{\parallel}$ and with $E_{\parallel} = - \nabla_{\parallel} \phi - \frac{1}{c} \frac{\partial}{\partial t} A_{\parallel}$, wherein we set $\phi = 0$ self consistently.
We linearise this equation and keep temperature fluctuations, in comparison to the original work. For describing the temperature fluctuations, we take the electron temperature equation and keep only the time derivative of $T_e$ and the parallel divergence of the parallel heat flux

\begin{equation}
    \frac{3}{2} \frac{\partial}{\partial t} T_e = - \frac{1}{n} \nabla \cdot \left( q_{\parallel e} \mathbf{b} \right) \, ,
    \label{eq:temperature_evolution}
\end{equation}

with $q_{\parallel e}$ the parallel conductive electron heat flux. All other terms vanish because of geometry or because they are higher-order terms. Linearising the parallel gradient of electron temperature leads to

\begin{equation}
    \nabla_{\parallel} T_e = \left( \nabla_z + \frac{i k_y \tilde{A}_{\parallel}}{B_0} \frac{\partial}{\partial x} \right) T_e = i k_z \tilde{T}_e + \frac{i k_y \tilde{A}_{\parallel}}{B_0} T^{\prime}_{e,0} \, ,
    \label{eq:grad_par_T_e}
\end{equation}

where a prime denotes the derivative in $x$-direction, $B_0$ is the background magnetic field and the fluctuating magnetic potential $\tilde{A}_{\parallel} = \exp(i k_y y + i k_z z - i \omega t)$ leads to $\tilde{B} = - \hat{z} \times \nabla \tilde{A}_{\parallel}$, $E_{\parallel} = - i \omega \tilde{A}_{\parallel}$ and $j_{\parallel} = - c/(4 \pi) \nabla_{\perp}^2 \tilde{A}_{\parallel}$, with $c$ the speed of light. Here, only the background derivative of the fluctuating temperatures and the flutter derivative of the background temperature survive. $\nabla_z T_{e,0}$ vanishes because the background temperature has no $z$-dependency and $i k_y \tilde{A}_{\parallel} \tilde{T}^{\prime}_e / B_0$ vanishes since it is a non-linear term.
We investigate both closures separately, by inserting $q_{\parallel e}$ in \cref{eq:temperature_evolution} according to the Braginskii or the Landau-fluid closure.

We start with the Landau-fluid closure, which is originally formulated in k-space already. The parallel wave vector $k_{\parallel}$ has a background component $k_z$ and a flutter component (see \cref{eq:grad_par_T_e}). The closure takes the form

\begin{equation}
    q_{\parallel e} = -A \frac{i k_{\parallel}}{|k_{\parallel}|} T_e \approx -A \frac{i k_z + ( i k_y \tilde{A}_{\parallel} / B_0 ) \, \partial_x}{|k_z|} T_e 
\end{equation}

with $A = n v_{e,th} \sqrt{8 / \pi}$ and $v_{e,th}$ the electron thermal velocity. The approximation on the denominator ($|k_{\parallel}| \approx |k_z|$), was also used for the simulation in \cref{sec:Turb_Sim}. Experience showed that enabling this term usually does not change the physical results of the simulation but increases computational cost, as also discussed previously \cite{pitzal_landau_2023} (sec. IV a). Inserting the Landau-fluid closure to \cref{eq:temperature_evolution} leads to

\begin{equation}
	-\frac{3}{2} \frac{i n \omega}{A} \tilde{T}_e = -\frac{k_z^2}{|k_z|} \tilde{T}_e - 2 \frac{k_z}{|k_z|} \frac{k_y \tilde{A}_{\parallel}}{B_0} T_{e,0}^{\prime} \, .
\end{equation}

Since there are still terms containing $|k_z|$, we investigate two cases $k_z \geq 0$ and $k_z < 0$. The first case leads to a negative growth rate in the end. Therefore, we continue with the second case ($k_z < 0$), which leads to 

\begin{equation}
\begin{split}
    \tilde{T}_e &= 2 \frac{k_y \tilde{A}_{\parallel}}{k_z B_0} T^{\prime}_{e,0} \; / \left( 1 - \frac{3}{2} \frac{i n \omega}{A k_z} \right) \\
    &\approx 2 \frac{k_y \tilde{A}_{\parallel}}{k_z B_0} T^{\prime}_{e,0} \left( 1 + \frac{3}{2} \frac{i n \omega}{A k_z} \right) \, .
\end{split}
\end{equation}

This approximation was checked for the observed temperature $T_e = 570 \; \mathrm{eV}$, density $n = 2.7 \cdot 10^{19} \, \mathrm{m^{-3}}$ and $k_y \rho_s = 0.2$ before the second PRE mentioned in \cref{sec:Global_PRE_mechanism} and assuming $a/L_n = a/L_{Te} = 15$ and $k_z = 8/R_0$ with $R_0 = 1.65 \, \mathrm{m}$, the minimal parallel wave number we can resolve with 16 poloidal planes. Since the real part of omega $\omega_r$ is much larger than the growth rate $\gamma$ and $\omega_r \approx \omega_p^{\star}$ for MTMs, we can show $\omega \approx \omega_p^{\star} \ll A k_z / n \approx \omega_p^{\star}/100$. Ohm's law now takes the form

\begin{equation}
    \begin{split}
        \eta j_{\parallel} &= E_{\parallel} + \frac{1}{en} \nabla_{\parallel} \left(p_{e,0} + \tilde{p}_e \right) + 0.71 \frac{1}{e} \nabla_{\parallel} \left( T_{e,0} + \tilde{T}_e \right) \\
        &= E_{\parallel} - i \tilde{A}_{\parallel} \left( \omega_p^{\star} + 4.13 \omega_T^{\star} \left( 1 + \frac{1.71}{4.13} \frac{3 n i \omega}{A k_z} \right) \right) 
    \end{split}
\end{equation}

with $\omega_p^{\star}$ and $\omega_T^{\star}$ as defined for \cref{eq:disp_rel_hatch}. Since the equation has the same form as in the original analysis, we find the same solution for the equation, only with slightly different parameters. The growth rate we end up with reads 

\begin{equation}
    \gamma = 1.71 \frac{3n}{A k_z} \omega_T^{\star} \left( \omega_p^{\star} + 4.13 \omega_T^{\star} \right) - \frac{\eta}{4 \pi} k_y^2 + ...
    \label{eq:lin_fluid_disp_rel}
\end{equation}

which provides a positive growth rate, if the first term exceeds the second one in magnitude. Exactly this growth rate is plotted in \cref{fig:path_in_grad_length_space} for the described values of $T_{e}$, $n$ and $k_y \rho_s$.

For the Braginskii closure 

\begin{equation}
    q_{\parallel e} = - \kappa_{\parallel e} \nabla_{\parallel} T_e \, ,    
\end{equation}

with $\kappa_{\parallel e}$, the parallel electron heat conductivity according to Braginskii \cite{Braginskii_1965}, \cref{eq:temperature_evolution} leads to an equation for the temperature fluctuations of the form 

\begin{equation}
	\tilde{T}_e = - 2 \frac{k_y \tilde{A}_{\parallel}}{k_z B_0} T_{e,0}^{\prime} \; / \; \left( 1 - \frac{3}{2} \frac{i n \omega}{k_z^2 \kappa_{\parallel e}} \right) \, .
\end{equation}

with the parallel heat conductivity $\kappa_{\parallel e}$ assumed to be constant. Following the same calculation as for the Landau-fluid closure leads to a term in the dispersion relation providing a negative growth rate, similar to the first case ($k_z > 0$) of the Landau-fluid closure. The idea that non-local effects are important for describing this kind of instability in this simplified geometry \cite{Hassam_1980, Guttenfelder_PRL_2011} is supported by our findings, since the $k_z / |k_z|$ is the term introducing non-local effects by the Landau-fluid closure and the one providing a positive growth rate in \cref{eq:lin_fluid_disp_rel}.

\section{\label{sec:Discussion}Discussion}

An important point is that although PREs are present in some I-mode discharges, discharge \# 37980 did not show any PREs at the relevant point in time. This discrepancy is worth discussing. As mentioned earlier, PREs are present especially in I-modes close to the I-H transition. This suggests that our fluid model predicts the appearance of MTMs and, thus, also PREs, for parameters that are still stable in the experiment.
A possible explanation lies in the resistivity $\eta$. In our simulations, we use the Spitzer resistivity for $\eta$, which has in SI units the form

\begin{equation}
    \eta_{\rm Sp} = \frac{4\sqrt{2\pi}}{3}\frac{Ze^{2}m_\text{e}^{1/2}\ln \Lambda}{\left(4\pi\varepsilon_0\right)^2 \left(k_\text{B}T_\text{e}\right)^{3/2}} \sim T_e^{-3/2},
\end{equation}

with $Z$ the charge number, $m_e$ the electron mass, $\ln \Lambda$ the coulomb logarithm, $\varepsilon_0$ the vacuum permeability and $k_B$ the Boltzmann constant. The important part is that $\eta_{Sp} \sim T_e^{-3/2}$, predicting very low values of resistivity for high temperatures. This discrepancy is not as tremendous as for the parallel heat fluxes, with a dependency of $q_{\parallel e}^{\mathrm{BR}} \sim T_e^{5/2}$ in the Braginskii closure, but it can still lead to issues for high temperatures. The fact that the parallel Spitzer resistivity underestimates the resistivity for high temperatures is commonly known \cite{Xu_PRL_2010} and was recently investigated quantitatively by comparing kinetic simulation results to values proposed by the Spitzer formula \cite{Tummel_2020}.  

Following this line of argument would imply that the value of $\eta$ should be larger in our simulations to come closer to reality. A larger value of $\eta$ would provide stronger damping of the linear growth rate due to the second term in \cref{eq:lin_fluid_disp_rel}, which would keep MTMs stable for the gradient lengths and temperatures present in our simulation and would be consistent again with observations in the experiment.

The MTMs observed in the simulation could therefore be thought of as artificially enhanced MTM, which occurs at a position where the electron temperature is high enough to significantly underestimate the resistivity and therefore also the damping part in the linear growth rate \cref{eq:lin_fluid_disp_rel}. This also explains why we observe our MTM not at the region of steepest gradients, more precisely of highest $\omega_p^{\star}$ as expected \cite{Hassan_2022}, but at a region with higher temperature and shallower gradients.

Although PREs are observed in our simulations earlier than in the experiment, the qualitative picture of PREs sketched in this paper should be unaffected.

Additionally, power-driven simulations were performed, with a fixed input power of $P_{\mathrm{in}} = 1.9 \; \mathrm{MW}$ equally split between electrons and ions, PREs were also present in these simulations. However, the electron temperature gradient steepened up close to the core boundary with the temperature value at the core boundary rising. In this case, the PRE occurred in this steep gradient region, lowering the temperature value at the core boundary due to the increased radial electron heat transport. Because the PRE appeared in the vicinity of the source region, the analysis would be a lot more questionable. Therefore, we focused on simulations with constant temperature values at the core boundary for this paper. 

\section{\label{sec:Summary_and_Outlook}Summary and Outlook}

We presented global trans-collisional plasma fluid simulations performed with \grillix, which showed intermittent events that are linked to PREs. The simulation was able to reproduce important features of the experimentally observed PREs. Furthermore, the underlying microinstability triggering PREs in our simulation was pinpointed by detailed analysis to MTMs. Since MTMs are only driven by the electron temperature gradient, this is an important finding, which might help to mitigate PREs in experimental discharges. 

The PREs observed in our simulation reproduce a wide range of experimentally observed features, e.g. increased magnetic activity, predominantly radial electron heat transport and the duration of the PRE matches. Therefore, we are confident that the presented qualitative picture for PREs is correct.
By comparing the path in gradient length space of the system to the normalised linear growth rate $\hat{\gamma}$ of the fluid model, we could identify and describe the global mechanism of a PRE cycle in our simulation, which is expected to be the same in the experiment.
Linear analysis in slab geometry revealed that the Landau-fluid closure with its non-local $k_{\parallel} / |k_{\parallel}|$ term predicts a positive growth rate in comparison to the Braginskii closure. Consistently, we did not observe PREs in the Braginskii simulation. The linear growth rate calculated for the Landau-fluid closure is in excellent agreement with the non-linear simulation and provides the basis for identifying the global PRE cycle. Furthermore, this linear growth rate points towards the Spitzer resistivity as a reason for overestimating the onset of PREs in our simulation.

The fluid model employed in this study shows limitations for low collisionality, not only but also due to the resistivity model. For future investigations in the I-mode regime, either for steady-state or addressing PREs, the model for the resistivity within \grillix~ should be improved \cite{Tummel_2020,Redl_2021} or models of higher fidelity like the gyrokinetic code \genex~\cite{Michels_genex_2021} should be used.
Combining the insights collected with \grillix~, which is very accurate in the SOL, but suffers in the confined region from the mentioned limitations of the fluid model, and future investigations with \genex~, which is very accurate in the confined region also for low collisionality, but suffers from missing sheath boundary conditions and neutral particle dynamics in the SOL, will hopefully shed some more light onto the physics underlying the I-mode regime.

\section*{Acknowledgements}

The authors would like to thank B.~Frei, G.~Merlo, D.~Silvagni, E.~Poli, P.~Manz and J.~Pfennig for fruitful discussions on many intricate topics. The simulations shown in this paper were conducted on the RAVEN Supercomputer hosted by the Max Planck Computing and Data Facility (MPCDF).

\section*{Bibliography}

\bibliographystyle{ieeetr}
\bibliography{paper.bib}

\end{document}